\documentclass[12pt]{article}

\begin{document}

\sloppy
\begin{flushright}{SIT-HEP/TM-18}
\end{flushright}
\vskip 1.5 truecm
\centerline{\large{\bf Curvaton paradigm can accommodate multiple low
inflation scales}} 
\vskip .75 truecm
\centerline{\bf Tomohiro Matsuda
\footnote{matsuda@sit.ac.jp}}
\vskip .4 truecm
\centerline {\it Laboratory of Physics, Saitama Institute of
 Technology,}
\centerline {\it Fusaiji, Okabe-machi, Saitama 369-0293, 
Japan}
\vskip 1. truecm
\makeatletter
\@addtoreset{equation}{section}
\def\theequation{\thesection.\arabic{equation}}
\makeatother
\vskip 1. truecm

\begin{abstract}
\hspace*{\parindent}
Recent arguments show that some curvaton field may generate the
cosmological curvature perturbation.
As the curvaton is independent of the inflaton field, 
there is a hope that the fine-tunings of inflation models 
can be cured by the curvaton scenario.
More recently, however, D.H.Lyth discussed that there is a strong
bound for the Hubble parameter during inflation 
even if one assumes the curvaton scenario.
Although the most serious constraint was evaded, the bound seems rather
crucial for many models of a low inflation scale. 
In this paper we try to remove the constraint.
We show that the bound is drastically modified if there were multiple
 stages of inflation. 
\end{abstract}

\newpage
\section{Introduction}
\hspace*{\parindent}
It has recently been proposed that the energy density perturbations
could originate in a scalar field other than the conventional inflaton
field\cite{curvaton_1,curvaton_2}, which is called curvaton.
After inflation, the curvaton starts to oscillate in a radiation
background. 
During this period the energy density of the curvaton grows to account 
for the cosmological curvature perturbation when it decays.
This curvaton paradigm has attracted a lot of attention because it has
an obvious advantage.
The inflaton energy scale is decoupled from the magnitude of the cosmic
microwave background(CMB) temperature fluctuations, which may cure
the fine-tunings of inflation models.
Along this line of thought, it seems attractive to suppose that the curvaton
paradigm can also save the models of a low inflation
scale\cite{curvaton_liberate}. 

The construction of a realistic model of low inflation scale is an
interesting problem especially when the fundamental scale is (much) lower than
the Planck scale, because in these 
models the Hubble parameter cannot become so large as the one in the
conventional models of inflation.  
In the history of the string theory, originally the inverse of the size
of extra dimensions as well as the 
fundamental scale was assumed to be as large as $M_p$. 
However, later observations showed that there is no reason
to require such a tiny compactification radius\cite{Extra_1}.
In models of large extra dimensions, 
the observed Planck mass is obtained by the relation $M_p^2=M^{n+2}_{*}V_n$,
where $M_{*}$ and $V_n$ denote the fundamental scale of gravity
and the volume of the $n$-dimensional compact space.
In the new scenario of the string theory, the compactification radius
and the fundamental scale are unknown parameters that should be
determined by observations.  
However, in many cases the cosmology of the large extra dimension
must be quite different from the conventional ones.\footnote{Finding crucial
constraints on the compactification, or constructing successful models
of the inflation with low 
(or intermediate) fundamental scale is a challenging
issue\cite{low_scale_inflation}.
In the models of large extra dimensions, the difficulties of inflation
are sometimes related to the stability of the compactified
space\cite{low_scale_inflation,stability} or the mechanism of
baryogenesis that must take place after
inflation\cite{low_baryo,low_AD,ADafterThermal}.} 

In spite of the above expectations, D.H.Lyth showed recently that there
 will be a strong bound for the Hubble parameter during inflation 
even if one assumes curvaton scenario\cite{Lyth_constraint}.
At the same time, a mechanism that evades the most serious
constraint is also suggested in ref.\cite{Lyth_constraint}.
However, to achieve the bound $H_I > 10 TeV$, a huge curvaton mass
 ($m_\sigma\sim M_p$) is required
 after the phase transition that takes place at the end of inflation.
Moreover, since the obtained bound is rather restrictive ($H_I> 10 TeV$), 
it should be fair to say that many models of a low inflation scale
are still not liberated by the curvaton\cite{curvaton_liberate}.

In this paper we show that the constraints obtained in
 \cite{Lyth_constraint} are relaxed to a satisfactory level
if there were multiple stages of inflation.
No large hierarchy between the scales of each inflation is required.
At least two stages of inflation are required in our scenario.
Denoting the vacuum energy during the two kinds of inflation by $V_1$ and
$V_2$, the original constraint is reduced by the factor of $\epsilon^6$, where
$\epsilon$ is the ratio of the scales, $\epsilon\equiv
(V_2/V_1)^{\frac{1}{4}}$.
As a result, for example, if the original bound for single inflation is
 $H_I>10^7 GeV$, 
we can reduce it as $H_1 >10^{-5} GeV$ for $\epsilon=10^{-2}$, where
$H_1$ is the Hubble parameter during the first inflation.
No sensible bound is obtained for the Hubble parameter during the second 
inflation.

\section{Constraint on inflation scales}
\hspace*{\parindent}
Here it may be helpful to begin with the review of the discussion in
ref.\cite{Lyth_constraint}.
First we start with more general settings and try to reproduce the
constraint showing how the original model in
\cite{Lyth_constraint} 
is realized in the boundary condition of our settings.
For simplicity, we assume that the curvaton field $\sigma$ is frozen
at $\sigma=\sigma_{osc}$ from the epoch of horizon exit during first
inflation to the epoch when 
the curvaton start to oscillate.
At the time when the curvaton starts to oscillate, its density is 
$\rho_\sigma \sim m_\sigma^2 \sigma_{osc}^2$.
Denoting the total density of other fields by $\rho_{tot}$, the ratio
$r$ at the time of curvaton oscillation is
\begin{equation}
\left.\frac{\rho_{\sigma}}{\rho_{tot}}\right|_{H=H_{osc}}
\sim \frac{m_{\sigma}^2 \sigma_{osc}^2}{H_{osc}^2 M_p^2}.
\end{equation}
During the period when the total density $\rho_{tot}$ is 
radiation-dominated, the ratio $r$ grows and reaches finally at
\begin{equation}
\label{rfirst}
r \le \frac{\sqrt{H_{osc} M_p}}{T_d }
\frac{m_\sigma^2 \sigma_{osc}^2}{H_{osc}^2 M_p^2}
\end{equation}
when the curvaton decay.
Here $T_d$ is the temperature just after curvaton decay.
Now the curvature perturbation 
\begin{equation}
\zeta \simeq \frac{r}{3}\frac{\delta \rho_{\sigma} }{\rho_\sigma}
\end{equation}
is generated by the curvaton.
Using the spectrum of the perturbation $<\delta
\sigma_{osc}^2>=(\frac{H_I}{2\pi})^2$,
the spectrum of the curvature perturbation is given by
\begin{equation}
\label{Pzeta}
{\cal P}_{\zeta}^{\frac{1}{2}}\simeq 
\frac{2r}{3}\frac{H_I}{2\pi\sigma_{osc}}.
\end{equation}
Using the required value from the observations 
${\cal P}_{\zeta}^{\frac{1}{2}}=5\times 10^{-5}$, one obtains that
\begin{equation}
\label{pert1}
\frac{2r}{3}\frac{H_I}{2\pi\sigma_{osc}}\simeq 5\times 10^{-5}.
\end{equation}
Using (\ref{rfirst}) and (\ref{pert1}), one finds the following
constraint;
\begin{equation}
\label{start}
\frac{2}{3}\frac{H_I}{2\pi\sigma_{osc}}
 \frac{\sqrt{H_{osc} M_p}}{T_d }
\frac{m_\sigma^2 \sigma_{osc}^2}{H_{osc}^2 M_p^2}
\ge 5\times 10^{-5}.
\end{equation}
Since the naive bound from nucleosynthesis is $T_d >1 MeV$,
one can obtain 
\begin{equation}
\label{constraint1}
\frac{2}{3}
\frac{H_I \sigma_{osc} m_{\sigma}^2}
{ 2\pi H_{osc}^{\frac{3}{2}} M_p ^{\frac{5}{2}}}
\ge 5\times 10^{-26}.
\end{equation}
On the other hand, one can use the lower bound for the curvaton decay
rate $\Gamma_{\sigma}\ge \frac{m_{\sigma}^3}{M_p^2}$ to obtain 
$T_d \simeq \sqrt{M_p \Gamma}\ge M_p (m_{\sigma}/M_p)^{\frac{3}{2}}$,
which implies that 
\begin{equation}
\label{constraint2}
\frac{2}{3}
\frac{H_I \sigma_{osc} m_{\sigma}^\frac{1}{2}}
{ 2\pi H_{osc}^{\frac{3}{2}} M_p }
\ge 5\times 10^{-5}. 
\end{equation}

The four parameters ($H_I, \sigma_{osc}, H_{osc},
m_{\sigma}$), are the boundary condition that depends on which model
one may choose.
Here the Hubble parameter during inflation when the observable
University leaves the horizon is denoted by $H_I$.
Considering (\ref{pert1}) and $r<1$, one finds
\begin{equation}
\label{ineq1}
\frac{2}{3}\frac{H_I}{2\pi \sigma_{osc}}
> 5\times 10^{-5}.
\end{equation}
From eq.(\ref{ineq1}) and (\ref{constraint1}), one finds
\begin{equation}
\label{constraint1a}
H_I > 10^{-15} \times 
\frac{H_{osc}^{\frac{3}{4}} M_p^{\frac{5}{4}}}{m_\sigma}.
\end{equation}
On the other hand, from eq.(\ref{ineq1}) and (\ref{constraint2}) one
obtains the following constraint;
\begin{equation}
\label{constraint2a}
H_I > 10^{-4} \times 
\frac{H_{osc}^{\frac{3}{4}} M_p^\frac{1}{2}}{m_\sigma^{\frac{1}{4}}}.
\end{equation}

First let us consider a simple example of the boundary condition.
We will assume that the mass of the curvaton $m_\sigma$ is a constant
during the period we are interested in.
Then the oscillation of the curvaton field starts when the Hubble
parameter falls below the curvaton mass, $H< m_\sigma$, which means
$H_{osc} \simeq m_\sigma$. 
As the inequality $H_I>H_{osc}$ always holds, we set a new parameter
$\epsilon_H$, which is defined as $H_{osc} \equiv \epsilon_H^2 H_{I}$.
Now one can replace $H_{osc}$ and $m_\sigma$ in eq.(\ref{constraint1a})
by $H_I$ to find 
\begin{equation}
\label{constraint1b}
H_I > M_p \times 10^{-12} \times \epsilon_H^{-\frac{2}{5}} 
\simeq 10^{6} \epsilon_H^{-\frac{2}{5}}  GeV.
\end{equation}
Following the same arguments, one can find from (\ref{constraint2a}),
\begin{equation}
\label{constraint2b}
H_I > 10^{-8} M_p \epsilon_H^2
\simeq 10^{10} \epsilon_H^2 GeV 
\end{equation}
As is discussed in ref.\cite{Lyth_constraint}, the bound from
eq.(\ref{constraint2b}) becomes more strict than (\ref{constraint1b})
when $\epsilon_H > 10^{-2}$.
Note that a tiny $\epsilon_H$ does not relax the bound.
The obtained bound $H_I >10^7 GeV$ corresponds to the
first constraint that was obtained in ref.\cite{Lyth_constraint}.

In the above arguments we showed that the bound obtained in
ref.\cite{Lyth_constraint} is derived from an initial condition that
fixes the four parameters ($H_I, \sigma_{osc}, H_{osc},
m_{\sigma}$).
Because some part of the initial condition is still model dependent,
it is interesting to find models that evade the above constraint
without introducing fine-tunings or large hierarchy.
A way to relax the bound is already suggested in \cite{Lyth_constraint}.
We think it is instructive to reproduce the argument within our setups and
show explicitly how the initial condition is modified in the model.
In the ``heavy curvaton'' scenario,
two differences appear in the initial condition.
Assuming that there was a phase transition just after inflation, which
makes the curvaton mass much larger than the one during inflation,
the conditions are modified as follows;
\begin{eqnarray}
H_{osc}\simeq m_{\sigma} \, \, &\rightarrow& \, \,  
H_{osc} \ll m_{\sigma}\nonumber\\
H_{osc}<H_I \, \, &\rightarrow& \, \, H_{osc} \simeq H_I.
\end{eqnarray}
Using the modified initial conditions and (\ref{constraint1a}), one can
easily find  
\begin{equation}
\label{constraint1c}
H_I > 10^{-56} \frac{M_p^5}{m_{\sigma}^4}.
\end{equation}
One can also find another bound from (\ref{constraint2a}),
\begin{equation}
\label{constraint2c}
H_I > 10^{-14}\frac{M_p^2}{m_\sigma}.
\end{equation}
As is suggested in \cite{Lyth_constraint}, it might be possible to obtain 
a preferable bound $H_I> O(TeV)$ if the curvaton mass becomes as
large as the Planck mass.
A huge curvaton mass may be allowed in a conventional supergravity,
but is not preferable in the models of large or intermediate extra dimensions.
Thus it is still an interesting problem to find models in which 
the bound is more relaxed.

\section{Multiple inflation}
\hspace*{\parindent}
In the followings we try to find another model whose initial condition
significantly 
lowers the above bound for $H_I$.
The most promising example will be to assume that the phase
transition that induces the curvaton oscillation starts independently of
the inflation that produces the spectrum of the perturbation $\delta
\sigma$.
The easiest way to realize the model is to introduce secondary
weak inflation that triggers the required phase transition.\footnote{
In this paper we do not mention how to realize the multiple
inflation.
Some arguments are given in \cite{low_scale_inflation,thermal}.}

Here we introduce the parameter $H_e$ that denotes the Hubble parameter during
the last inflation. 
The spectrum of the perturbation $\delta \sigma$ is produced during the
first inflation, the Hubble parameter during which is denoted by $H_I$.
More explicitly, the parameters follow the conditions, $H_I \gg H_e$,
$m_\sigma \gg H_{osc}$ and $H_{osc}\simeq H_e$.
The bound (\ref{constraint1a}) now becomes 
\begin{equation}
\label{constraint1d}
H_I > 10^{-15} \frac{H_{e}^\frac{3}{4} M_p^\frac{5}{4}}{m_\sigma}.
\end{equation}
Of course, setting $H_I=H_e=H_{osc}$, one obtains (\ref{constraint1c})
again.
The bound (\ref{constraint2a}) becomes
\begin{equation}
\label{constraint2d}
H_I > 10^{-4} \frac{H_{e}^\frac{3}{4} 
M_p^\frac{1}{2}}{m_\sigma^{\frac{1}{4}}}.
\end{equation}
Now let us consider a case when $H_{e}=H_I \times
10^{-4}$.
In this case, (\ref{constraint2d}) becomes
\begin{equation}
\label{constraint2d2}
H_I > 10^{-26} \frac{M_p^2}{m_\sigma}
\end{equation}
which is about $10^{-12}$ times smaller than (\ref{constraint2c}).
To see exactly what happened in the bound, here we consider a more
generic situation.
Introducing mass scales $M_I \equiv V_I^{\frac{1}{4}}=H_I^2M_p^2$ and
$M_e \equiv V_e^{\frac{1}{4}}=H_e^2M_p^2$, and denoting the ratio by
$\epsilon \equiv \frac{M_e}{M_I}$, one finds from (\ref{constraint1d}),
\begin{equation}
\label{constraint1e}
H_I > 10^{-56} \epsilon^6 \frac{M_p^5}{m_\sigma^4}.
\end{equation}
One can also find from (\ref{constraint2d}),
\begin{equation}
\label{constraint2e}
H_I > 10^{-14} \epsilon^6 \frac{M_p^2}{m_\sigma}. 
\end{equation}
From eq.(\ref{constraint1e}) and eq.(\ref{constraint2e}), it is easy to
understand why the bound is so sensitive to the ratio $\epsilon$.
If the scale of the vacuum energy during the first inflation is 
only $10^{2}$ times larger than the one during the second inflation,
the bound is significantly reduced by the factor of $(10^{-2})^6$.

\section{Conclusions and Discussions}
\hspace*{\parindent}
In this paper we have examined the constraint on the Hubble parameter
during inflation when the curvaton hypothesis is used to explain the
observed density perturbation of the Universe.
The constraint (\ref{start}) that we have started from
produces bounds for the Hubble parameter once the initial conditions are  
determined by the models.
We have found an example of the model that significantly modifies the
bound.
At least two stages of inflation are required in our scenario.
The first produces the
spectrum of the perturbation $\delta \sigma$ while the other triggers
the phase transition that produces the curvaton mass.
Then the ratio of the vacuum energy $\epsilon$ reduces the constraint by the
factor of $\epsilon^6$, which evades the bound that was
discussed in \cite{Lyth_constraint}.

\section{Acknowledgment}
We wish to thank K.Shima for encouragement, and our colleagues in
Tokyo University for their kind hospitality.

\end{document}